\documentclass[12pt]{article}
\usepackage{times}
\usepackage{wrapfig}
\usepackage{geometry}
\usepackage{graphicx}
\usepackage[utf8]{inputenc}
\usepackage{enumitem,amssymb}
\usepackage{ragged2e}
\usepackage{pifont}

\def\iso#1#2{\mbox{${}^{#2}{\rm #1}$}}
\def\fe6#1{\iso{Fe}{6#1}}
\def\pu24#1{\iso{Pu}{24#1}}
\def\al2#1{\iso{Al}{2#1}}

\def\ga{\stackrel{>}{\sim}}

\def\bul{$\bullet$}

\def\citeme#1{\citep{#1}}

\usepackage[numbers]{natbib}

\def\apj{ApJ}
\def\apjl{ApJL}
\def\nat{Nature}
\def\araa{Annual Reviews of Astronomy and Astrophysics}
\def\ssr{Space Science Reviews}
\def\aap{Astron.~and Astrophys.}
\def\prd{Phys.~Rev.~D}

\begin{document}
\raggedright
\LARGE
    {\sffamily \bfseries Astro2020 Science White Paper} \linebreak
    \bigskip

{\sffamily \bfseries Near-Earth Supernova Explosions:  \\
Evidence, Implications, and Opportunities} \linebreak
\normalsize

\bigskip

{\sffamily \bfseries \large Submitted to: \\
  The 2020 Decadal Survey on Astronomy and Astrophysics \\
  U.S. National Academies of Sciences, Engineering, and Medicine \\
  \smallskip
  Committee on Astronomy and Astrophysics}

\vspace{0.5in}

\noindent {\sffamily \bfseries  Thematic Areas:}

\smallskip
%\begin{tabular}{l}
  $\square$ Stars and Stellar Evolution  \\
  $\square$ Resolved Stellar Populations and their Environments \\
  $\square$ Formation and Evolution of Compact Objects  \\
  $\square$  Multi-Messenger Astronomy and Astrophysics
%  \end{tabular}
%~~\\

\newpage
 
{\sffamily \bfseries Authors:} \bigskip

Brian~D.~Fields, University of Illinois. {\em (Corresponding author)} \\
John~R.~Ellis, King's College London \\
Walter R.~Binns, Washington University in St.~Louis \\
Dieter Breitschwerdt,  Berlin Institute of Technology  \\
Georgia A. deNolfo, Goddard Space Flight Center \\
Roland Diehl, Max Planck Institut f\"ur Extraterrestrische Physik \\
Vikram V.~Dwarkadas, University of Chicago \\
Adrienne~Ertel, University of Illinois \\
Thomas Faestermann, Technische Universit\"at M\"unchen \\ 
Jenny Feige,  Berlin Institute of Technology  \\
Caroline~Fitoussi, \'Ecole Normale Sup\'erieure, Lyon \\
Priscilla Frisch, University of Chicago \\
David Graham, Oregon State University \\
Brian Haley, Oregon State University \\
Alexander~Heger, Monash University \\
Wolfgang Hillebrandt, Max-Planck-Institut f\"ur Astrophysik \\
Martin H.~Israel, Washington University in St.~Louis \\
Thomas Janka, Max-Planck-Institut f\"ur Astrophysik \\
Michael Kachelrei{\ss}, Norwegian University of Science \& Technology, Trondheim \\
Gunther Korschinek, Technische Universit\"at M\"unchen \\ 
Marco Limongi, INAF/Osservatorio Astronomico di Roma, KIPMU/Tokyo \\
Maria Lugaro, Konkoly Observatory, Hungarian Academy of Sciences,
and Monash University \\
%Eric~E.~Mamajek, Jet Propulsion Laboratory \\
Franciole~Marinho, Universidade Federal de S\~ao Carlos \\
Adrian~Melott, University of Kansas \\
Richard A.~Mewaldt, California Institute of Technology \\
Jesse~Miller, University of Illinois \\
Ryan C.~Ogliore, Washington University in St.~Louis \\
Michael Paul, Hebrew University \\
Laura Paulucci, Universidade Federal do ABC \\
Mark Pecaut, Rockhurst University \\
Brian F.~Rauch, Washington University in St.~Louis \\
Karl E.~Rehm, Argonne National Laboratory \\
Michael Schulreich, Berlin Institute of Technology \\
Ivo Seitenzahl, University of New South Wales and Australia National University \\
Mads S{\o}rensen, University of Geneva \\
Friedrich-Karl Thielemann, University of Basel and GSI Darmstadt \\
Francis X.~Timmes, Arizona State University \\
Brian C.~Thomas, Washburn University \\
Anton Wallner, Australia National University 
%\linebreak

%\vspace{-4mm}

\pagebreak

{\bf \large Executive Summary}

{\bf There is now solid experimental evidence of at least one supernova explosion within 100~pc of Earth
within the last few million years}, from measurements of the short-lived isotope $^{60}$Fe in widespread
deep-ocean samples, as well as in the lunar regolith and cosmic rays. This is the first established example of a specific dated
astrophysical event outside the Solar System having a measurable impact on the Earth, offering new probes of
stellar evolution, nuclear astrophysics, the astrophysics of
the solar neighborhood, cosmic-ray sources and acceleration,
multi-messenger astronomy, and astrobiology.
Interdisciplinary connections reach broadly to include heliophysics, geology, and
evolutionary biology. Objectives for the future
include pinning down the nature and location of the established near-Earth supernova explosions, seeking evidence for others,
and searching for other short-lived isotopes such as $^{26}$Al and $^{244}$Pu.
{\bf The unique information provided by geological and lunar detections of radioactive \fe60
to assess nearby supernova explosions make now a compelling time for the astronomy community
to advocate for supporting multi-disciplinary, cross-cutting research programs.}
\\ \bigskip

{\bf \large Geological and Lunar Detections of Radioactive Iron-60 
 as Evidence for Near-Earth Supernovae} 
\begin{wrapfigure}[28]{l}{0.48\textwidth}
%\begin{figure}%[t]
%\includegraphics[width=7.4cm,height=5.5cm]{60fe_data_crusts_other.eps}\\
%\hspace*{2mm}
%\includegraphics[width=7.4cm,height=5.5cm]{60fe_data_sediments.eps}
\includegraphics[width=7.8cm]{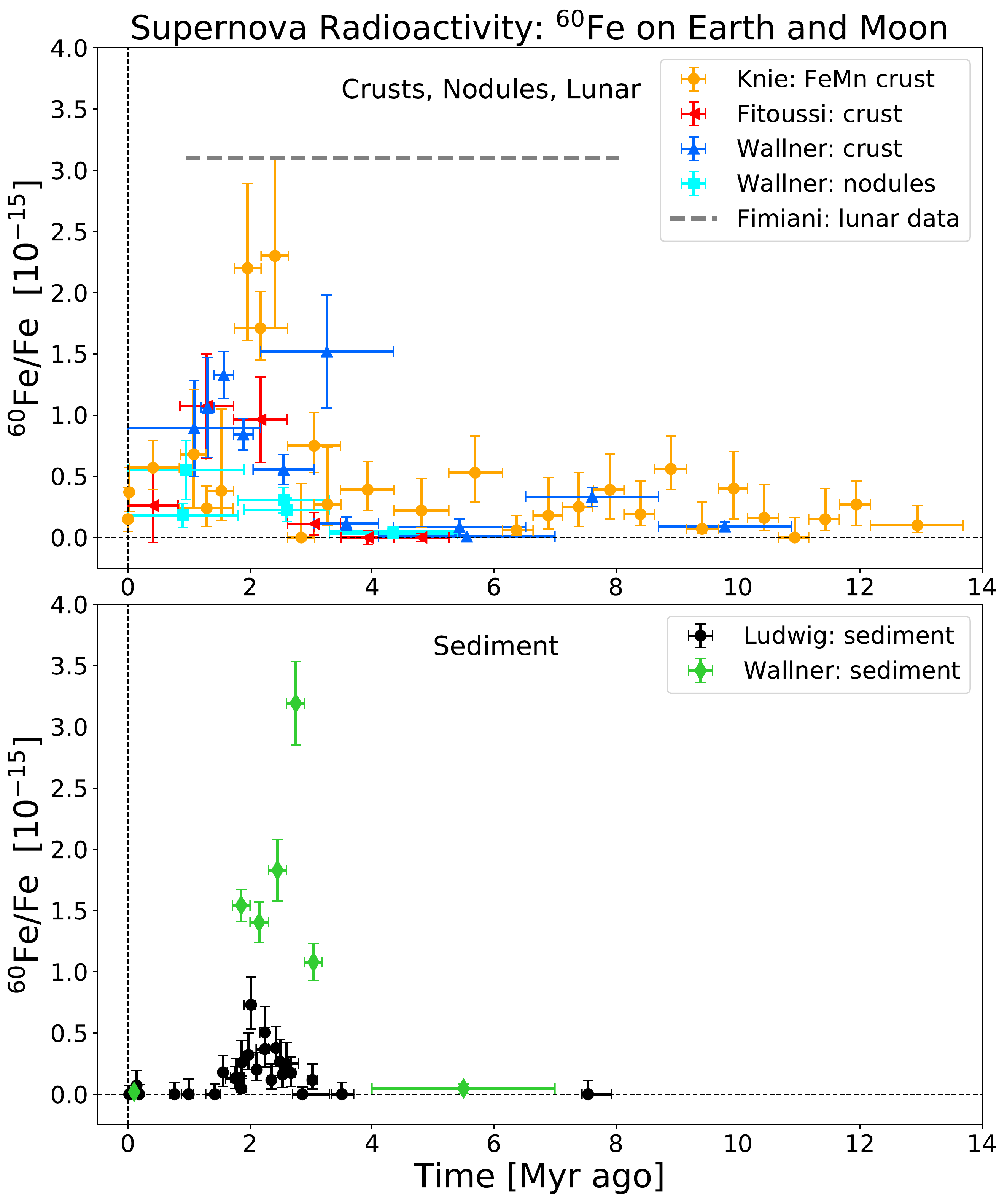}
\caption{\small Global and lunar detections of \fe60,
  not corrected for decay.
  All data show a signal around $\sim$2--3 Myr. 
  Amplitude differences may reflect 
  iron uptake variations, or latitude variations in iron fallout.
  {\em Upper panel:} $^{60}$Fe/Fe ratios in
  deep-ocean Fe-Mn crusts.  %There is a hint of an additional signal $\sim 8 \ \rm Myr$ ago.  
  %{\bf John: Is the right place for the lunatics?}
  {\em Lower panel:} $^{60}$Fe/Fe in deep-ocean sediments,
  %{\bf John: one `sediment' mis-spelt.}
  showing signal duration $\ga 1 \ \rm Myr$.
  %Note AMS sensitivity required to detect $\rm \fe60/Fe \sim 10^{-15}$.
  Data: refs.~\citep{knie2004,fitoussi2008,wallner2016,fimiani2016,ludwig2016}.%\\
%{\bf THESE WILL BE STACKED INTO ONE FIGURE.} 
\label{crusts+sediments}  }
%\vspace*{3mm}
\end{wrapfigure}
\quad Near-Earth~supernovae~are~inevitable. Supernovae~explode~in~our~Milky~Way roughly~every~$\sim$~30~yr~on~average~\citep{diehl2006,adams2013}. This~suggests~that~within~the~past billion years,~one~or~more~supernovae~may have~exploded~$\le$~10~pc~of~the~Earth,~with
drastic~effects~on~the~biosphere,~possibly
producing~a~mass~extinction~\citep{schindewolf1954,ruderman1974,gehrels2003,korschinek2017}.
Alvarez~et~al.~\citep{alvarez1980}~hypothesized~that~such
a~nearby~supernova~explosion~would~have
deposited~a~detectable~live~(not~decayed)
radioisotope~layer~on~Earth.~Searching~for
it,~they~found~the~iridium layer~near~the~K/Pg
boundary,~associated~instead~with~a~bolide
impact~responsible~for~the~demise~of~the~dinosaurs. \\
%but they crucially identified radioisotopes
%as the smoking gun of a near-Earth supernova.
\quad
Analogously,~supernova~explosions~within
100~pc~of~Earth~are~expected~to~have~occurred
every~few~Myr.~The~Local~Bubble~surrounding
the~Sun~implies~nearby~events~within~2~Myr~\cite{frisch1981}.
These~would probably~not~have~caused~a~mass
extinction,~but~may~have~perturbed~the~biosphere
and~left~a~detectable~radioisotope~signature.
Ref.~\citeme{ellis1996}~suggested~several~possible
radioisotope~signatures~of~such~a~supernova,~including
$^{60}$Fe~(see also ref.~\cite{Korschinek1996}).\\
\quad
Using~the~spectacular~sensitivity~of
accelerator mass spectrometry (AMS), a layer of $^{60}$Fe (half-life 2.6 Myr) was discovered
in a  ferromanganese (FeMn) crust sample from the Pacific Ocean floor \citeme{knie1999}, and confirmed with a different
FeMn crust from elsewhere in the Pacific 
\citeme{knie2004}.
Subsequent FeMn crust measurements further confirmed these pioneering 
results \citeme{fitoussi2008}.  A first series of sediment samples from North
Atlantic revealed an \fe60 peak, but required a much longer
deposition time ($>0.4 \ \rm Myr$) than naively expected for supernovae
 %and a need to
%reconsider the uptake factor in the crust to significantly higher values
\citeme{fitoussi2008}.
This was later confirmed in Indian Ocean sediment samples \citeme{feige2014}.
Ref.~\citeme{wallner2016} also found an 
$^{60}$Fe signal in different Fe-Mn crusts and nodules, as well as in several deep-ocean sediments.
Moreover, ref.~\citeme{ludwig2016} detected $^{60}$Fe in iron-bearing microfossils found in deep-ocean sediments.  All sediment data reveal deposition timescales $\ga 1 \ \rm Myr$.

\vspace{-0.8cm}	
\hspace{6cm}
\begin{wrapfigure}[35]{l}{0.495\textwidth}
%\begin{figure}%[t]
%\centering
%\includegraphics[width=7.4cm,height=5.5cm]{Figure1.eps}
  %\includegraphics[width=7.5cm]{FFE_Drad.pdf}
  %\includegraphics[width=9.5cm]{Figure_Fluence.pdf}
  \includegraphics[width=8.25cm]{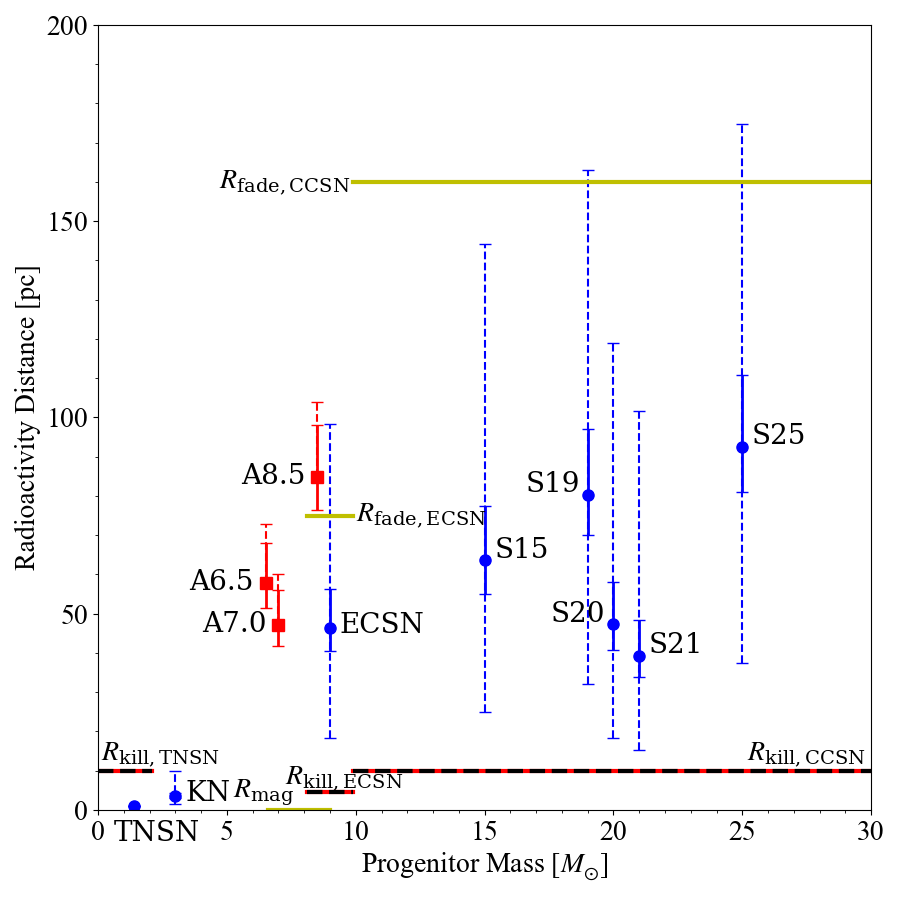}
%\hspace*{2mm}
%\includegraphics[width=7.4cm,height=5.5cm]{60fe_data_sediments.eps}
%
\caption{\small Distances to \fe60 sources estimated from
  \fe60 data and yields from progenitor models of
   various masses \citep{fry2015}.
  Distance $D \propto \sqrt{M_{\rm ej,60}/{\cal F}_{60}}$ from
    the inverse square law assuming: (1) each source spreads its
    \fe60 ejecta of mass
    $M_{\rm ej,60}$ isotropically, and (2) these lead to the observed
    deep-ocean \fe60 decay-corrected fluence ${\cal F}_{60}$.
    The horizontal lines bound possible distances:
    lower limit is the ``kill radius'' and excludes
    Type Ia or thermonuclear supernovae
    (TNSN) and kilonovae (KN).  
    Upper limit is where supernova remnant fades and fails to
    deliver \fe60 to Earth. The allowed sources based on distance are
    AGB stars (red, denoted A) and core-collapse supernovae (blue, denoted S).
    Of these only core-collapse events are likely to deliver the \fe60 to Earth;
    a wide range of masses are allowed, with distances $D_{\rm SN} \sim 30-150 \ \rm pc$.
\label{source}  }
%\vspace*{3mm}
\end{wrapfigure}

\quad
The~time~series~of~these~$^{60}$Fe measurements are shown in~Fig.~\ref{crusts+sediments}.~There is a clear
peak at 2 to 3 Myr ago, pointing to at least one nearby supernova at that epoch. Also,
there are hints of a second
peak around 8~Myr ago.

%~~\\

\quad
In~addition~to~these~time-resolved~terrestrial data,~ref.~\citep{fimiani2016}~report~an~$^{60}$Fe~excess~in~undated Apollo~return~samples~of~the~lunar~regolith. This~ties~in~with~the~discovery~of~$^{60}$Fe~in~cosmic
rays~\citeme{binns2016},~which~would~require~a~supernova
origin~in~the~last~$\sim$~2.6~Myr~within~$\sim$~1~kpc~of
Earth,~based~on~the~$^{60}$Fe~lifetime~and~models~of
cosmic-ray~diffusion.~This~is~consistent~with
studies~\citeme{kachelriess2015,savchenko2015,kachelriess2018}~that~used~cosmic-ray
spectra~to~argue~for~an~injection~of~cosmic~rays~by
a~supernova~occurring~$\sim 2$~Myr~ago. \\
%\hspace{-0.5cm}
\hspace{10cm}
       {\bf \large Interpretation}\\
%\quad
The production site of the observed $^{60}$Fe signal must have been a core-collapse supernova (CCSN).
For any~given~\fe60 nucleosynthesis source, supernova or other, one can use the measured fluence to estimate the distance to the progenitor. Fig.~\ref{source}
shows the result of one such analysis \citep{fry2015} that used models with a variety of progenitor masses for supernovae, but also considered \fe60 production
in AGB stars, Type Ia (thermonuclear) supernovae, and kilonovae.
The last two are ruled out by the implausibly and dangerously small
distance required by their small \fe60 yields.  AGB stars do exist 
at the allowed distances, but it is unlikely \fe60 from AGB
winds could reach Earth prior to decay.
CCSN emerge as the only viable candidates, and 
Fig.~\ref{source} shows CCSN models suggest a distance $\sim 100 \ \rm pc$, reassuringly beyond the possible `kill radius'
for a mass extinction, but still close enough to possibly
affect the biosphere \citeme{thomas2016,melott2017}.
We note in this connection the existence of young nearby pulsars that might be candidates for its compact remnant~\cite{tetzlaff2013}.\\
\quad
The firm \fe60 detection at $2-3$ Myr and possible additional signal at $\sim 8$ Myr
is consistent with the Sun's being surrounded by the
Local Bubble, the low-density, high-temperature
region that is thought to arise from a number of 
recent nearby supernova explosions.  The Local
Bubble model of ref.~\citeme{breitschwerdt2016} is
compatible not only with terrestrial and lunar detections of $^{60}$Fe
\citep{schulreich2017}, but also with two soft X-ray emitting cavities
present in the dust distribution of the local interstellar medium,
matching the sites of the two most recent supernova explosions in their
model with respect to both distance and direction: see
refs.~\citeme{capitanio2017,schulreich2018}. \\
\quad
Ref.~\citeme{ellis1996}  suggested~several~other radioisotopes besides $^{60}$Fe as possible signatures of a nearby supernova. Of these,
$^{26}$Al (half-life $0.73  \ \rm Myr$) and $^{244}$Pu (half-life
$81 \ \rm Myr$) have been the subjects of AMS searches in deep-ocean samples.
Recent $^{26}$Al data
are, however, compatible with terrigenic production,
which could point to low supernova yields \citeme{feige2018}
or to differences in Fe-bearing vs Al-bearing supernova dust \cite{fry2015}.
Intriguingly, ref.~\citeme{wallner2015} found $^{244}$Pu in the same
crusts as the \fe60 detections, but the results lie far
below what one would expect from continuous production in the Galaxy. This suggests that plutonium sources are rare, so that a steady-state equilibrium is not achieved over the \pu244 lifetime.  Rather, \pu244 sources are few and far between,
likely being neutron-star mergers \citeme{abbott2017}.
It remains, however, a critical test to probe more sensitively
the \pu244 signal {\em contemporary} with the \fe60 which will test whether the recent near-Earth event(s)
included {\rm r}-process production.  Confirming that some supernovae produce
{\em r}-process elements \citep{cowan2019,banerjee2018}, and quantifying the yields
would provide a major new probe of supernova
explosion physics and nucleosynthesis.
\\
\quad
An important issue in interpreting terrestrial signatures of a nearby supernova is modeling radioisotope transport from
the source to an ocean floor. Transport through the interstellar medium is thought to be via dust particles, which will
also interact with their natal supernova remnant and possibly
interstellar material. Important questions are the duration of this process, and whether the arrival directions of the 
dust particles are correlated with the direction of the source. The latter seems unlikely \citep{fry2018}, but could be tested with the lunar distribution
of radioisotopes.
On Earth,  atmospheric and oceanic effects are expected to destroy any information on the source latitude
and may cause variations in the density of deposition on the Earth's surface, 
concentrating $^{60}$Fe in certain areas of the ocean floor.\\
%~~\\
\bigskip

{\bf \large Possible Biological Effects}\\
\quad
The spectacular optical display of a nearby supernova
would not be very dangerous to life if the explosion is
$\sim 100 \ \rm pc$ away. The outburst would also bring higher-energy 
ionizing radiation, including extreme UV, X-rays and gamma rays,
yet even these are not catastrophically harmful 
from $\sim 100 \ \rm pc$.
However, charged cosmic rays
would arrive later with the supernova blast that accelerates them,
and linger for many thousands of years.
They would deplete the
Earth's ozone layer, which would in turn allow more solar UVB radiation to reach the Earth's surface and upper ocean layers
for an extended period~\citep{ellis+schramm1995,melott2018}.
Increases in ionizing radiation can 
damage DNA, harm tissues in animals, and degrade photosynthesis in plants.
Penetrating cosmic-ray muons may also be a hazard, as well as other 
effects of increased atmospheric ionization by cosmic rays \citep{melott2017,melott2019a,melott2019b}.
However, no `smoking gun' effect of a supernova at $\sim 100 \ \rm pc$ 
on the Earth's biosphere has yet been identified. A supernova at $10 \ \rm pc$ would surely be very dangerous for the biosphere,
but a distinctive signature remains to be found in the geological record.\\
~~\\

{\bf \large Outlook: Interdisciplinary Opportunities for the Coming Decade}

Following the large infusion of new data in 2016, this field has evolved from speculation and pioneering results to 
becoming a growing science, ripe with opportunity for surprise and discovery.
It has become a {\it bona fide} part of astrophysics, which is akin to and has many links with meteoritic studies,
particularly studies of extinct radioactivities in meteorites \citep{meyer2000,banerjee2016}
and pre-solar grains that seek to identify nucleosynthesis products from individual events \citep{clayton2004}.
%However, it is clear that in this case one is studying one or more very recent astrophysical events,
%specifically supernovae.
\\
\quad
{\bf Supernova Astrophysics:}
Detection of or strong limits on other radioisotopes
will be particularly useful in identifying the progenitor and potentially
its non-isotropic element production and dispersal.
Detection of {\em r}-process species would offer the first concrete evidence that supernovae produce these species.
Detection of the third-peak element \pu244
would require a major revision of our current
picture of the physics in typical core-collapse
events.     \\
\quad
{\bf Cosmic Dust:}
At distances $\ga 10 \ \rm pc$, the supernova blast does not reach
within 1 AU, instead radioactive debris rains on Earth in the
form of dust \citep{benitez2002,athanassiadou2011}.
The existing detection of \fe60, and future detections or limits to other
radioisotopes, thus directly probe the formation and evolution of
supernova dust.  This provides new insight into critical questions
about the formation and distribution of dust in our Galaxy as well
as supernovae at high redshift \citep{nozawa2007,gall2011}. \\
\quad {\bf Nuclear Astrophysics:}
The terrestrial detection of supernova debris offers a new laboratory probe
of element formation in the cosmos.
It is complementary to other aspects of astronomy with radioactivities, particularly gamma-line telescopes
that have detected \fe60 \cite{wang2007,diehl2018}
and mapped \al26 \cite{diehl2006}
in our Galaxy (see MeV line White Paper \citep{fryer2019}),
and the growing field of multi-messenger studies of kilonovae.
In particular, it casts direct experimental light on galactic nucleosynthesis, demonstrating the role of
supernovae in making specific isotopes \citep{tur2010,limongi2006,sukhbold2016}.  \\
\quad
{\bf Cosmic-Ray Astrophysics:} Local supernovae have been suggested as explanation for spectral anomalies
\citeme{kachelriess2015,savchenko2015,kachelriess2018}
and \fe60 detections \cite{binns2016} in cosmic rays.
Cosmic-ray disturbance of the biosphere and atmosphere can
lead to damage \cite{melott2017,melott2018} or even benefits \cite{knie2004,faestermann2018}.
Detailed study of cosmic rays
inside the Local Bubble and heliosphere
will be an important to test this scenario.
\\
\quad
    {\bf Solar Neighborhood:}
    Our Galactic environment depends
    on the frequency of nearby supernovae.
    The recent explosion(s) subject
models of the Local Bubble to new
experimental constraints \citep{frisch2011,breitschwerdt2016,frisch2017}.
Also, because stars are generally born in clusters, there is the opportunity
to identify the natal cluster of the \fe60 supernova(e).
Candidates include
the Scorpius-Centaurus association ($\sim 120 \ \rm pc$ away 3 Myr ago \citep{benitez2002})
and the Tucana-Horlogium association
($\sim 50 \ \rm pc$ away \citep{mamajek2016}),
and others \citep{hyde2018,sorensen2017}.
\\
\quad
    {\bf Solar System Formation:} Live radioactivities are known to have been
    present at the formation of the Solar System, likely implying a nearby supernova
    \cite{meyer2000,banerjee2016}.
    The commonalities between early-Solar and
    recent nearby supernova studies merit further exploration and connection \cite{looney2006,hotokezaka2015}.
     \\
\quad
    {\bf Heliophysics:}  The confirmation of at least one supernova
    $\sim 2-3 \ \rm Myr$ ago provides
    a unique opportunity to study the heliosphere
    under conditions dramatically different from the present \citep{fields2008,frisch2017}.
Supernova-driven shocks and radiation bursts play pivotal roles in regulating
the diffuse interstellar material around the heliosphere and nearby planetary
systems.
%Supernova shocks disrupt the kinematics of the ambient low-density interstellar medium.
Interstellar ram pressure determines the dimensions of
astrospheres \citep{frisch1993} and the heliosphere \citep{mueller2006}, and
so the supernova evolution sets the duration and closest approach
of the blast.
Neutral
interstellar atoms penetrate the heliosphere and mass-load the solar wind, see
refs.~\citeme{cummings2002,mccomas2009}, modulating the cosmic-ray flux
at Earth \citep{mueller2006} and regulating the boundary conditions of the
heliosphere \citep{slavin2008}. \\
\quad
{\bf Astrobiology and Planetary Science:}
%It opens up many issues in astrobiology, such as the possibility that o
Data on terrestrial effects of the recent explosion(s)
calibrates the impact of supernovae on the biosphere,
informing studies of a supernova-induced mass extinctions.
More generally, studies of the location and frequency of supernovae can cast light on which Milky Way regions are suitable for life--the Galactic Habitable Zone 
\citep{lineweaver2004,gonzalez2001,morrison2015}.\\
\quad {\bf Beyond Astronomy:  Across the Sciences}.
Near-Earth supernovae studies interweave
disparate sciences, including the following. {\em Nuclear Physics:}
This area of research has proven to be a novel and fruitful application of accelerator mass spectrometry (AMS) and its exquisite sensitivity (now pushing to levels of $\rm \fe60/Fe \sim 3 \times 10^{-17}$).
It is therefore encouraging that the pioneering AMS work by the Munich groups has recently been joined in these studies
by a group using an AMS facility in Australia.
{\em Geology:}  Geochronological and geochemical studies are crucial
for testing the atmospheric and climatic and even biological
impact of the event.  The established existence of
supernova(e) at the early Pleistocene offers a unique opportunity to study
the Earth's response to such a perturbation. \\
\quad
{\bf Open Questions and Future Research:}
There are many open questions that provide opportunities for future research.
What other radioisotopes were deposited on the Earth and Moon, and what
do these imply for supernova physics and supernova dust?  Can the type of supernova
responsible for the event $\sim 2.5$~Myr ago be identified, including the mass of its progenitor and the likely
nature of its ejecta? Can the location of the progenitor be identified, and could one even imagine identifying its
remnant?  Can one link this supernova to features in the Local Bubble? Were there other nearby supernovae 
within the past few million years? Can a link be established between any supernova event and some specific 
perturbation of the biosphere?\\
\bigskip

{\bf \large Enabling Discovery}\\
The time is therefore ripe to answer these questions through the combined efforts of the astrophysics
community and the other disciplines embraced by near-Earth supernova studies.   \\
\bul\ {\bf First and foremost,
we call for funding agencies to provide mechanisms to support cross-cutting
research of this kind.}  We emphasize our experience that the
novelty and interdisciplinarity of this subject are great strengths,
but pose challenges to funding within particular disciplines
(e.g., ``too geological for astrophysics, too astrophysical for geology'').
 \\
\bul\ {\bf We call for support of theoretical explorations of
the astrophysical aspects of this problem}, as individual areas of study
and in integrated syntheses.  These include supernova
nucleosynthesis, supernova dust formation, propagation and evolution,
cosmic-ray injection, acceleration and propagation, and supernova
impact on the heliosphere. \\
\bul\ {\bf We call for support of MeV gamma-ray observatories} capable of probing
nuclear lines from \fe60, \al26, and other supernova radioisotopes \citep{fryer2019}. \\
\bul\ {\bf We call for support by astrobiology programs} of studies of
the direct and indirect
impact of ionizing radiation (photons and cosmic rays)
on terrestrial biota, and of consequences for exoplanets and the Galactic habitable zone. \\
\bul\ {\bf We call for support of AMS}, not only to understand better the
$^{60}$Fe signal but also to measure or constrain additional radioisotopes, including known 
core-collapse supernova products such as $^{26}$Al \citep{feige2018}, but also other species probing other nucleosynthesis processes,
e.g., $^{182}$Hf and $^{244}$Pu \citep{wallner2015} that are key probes of the {\em r}-process, as well as other isotopes \citep{lugaro2018}.  \\
\bul\ {\bf We call for support of geoscience research in this area.}  Geochronologic and geochemical studies of terrestrial archives (e.g., of \iso{He}{3}) will
help to better constrain the timing/duration of the event(s),  as well as test their potential impact on changes in climate and even biological evolution. 
%{\bf John: Where did $^{182}$Hf, $^{97}$Tc and $^{98}$Tc jump from?}

%cites work the ApJ way:  in-text \citet{fry2015}, parentheses \citep{knie2004}

\pagebreak
%\textbf{References}


\begin{thebibliography}{abbott2017}

\bibitem[Abbott et al.(2017)]{abbott2017} Abbott, B., et al. \ 2017,  \apjl,  848,  L12

\bibitem[Adams et al.(2013)]{adams2013} Adams, S.~M., Kochanek, C.~S., Beacom, J.~F., Vagins, M.~R., \& Stanek, K.~Z.\ 2013, \apj, 778, 164 

\bibitem[Alvarez et al. (1980)]{alvarez1980} Alvarez, L., Alvarez, W., Asaro, F., \& Michel, H. 1980, Science, 208, 4448, 1095

\bibitem[Athanassiadou \& Fields(2011)]{athanassiadou2011} Athanassiadou, T., \& Fields, B.~D.\ 2011, New Astronomy, 16, 229 

\bibitem[Banerjee et al.(2016)]{banerjee2016} Banerjee, P., Qian, Y.-Z., Heger, A., \& Haxton, W.~C.\ 2016, Nature Communications, 7, 13639

\bibitem[Banerjee et al.(2018)]{banerjee2018} Banerjee, P., Qian, Y.-Z., \& Heger, A.\ 2018, \apj, 865, 120   
  
\bibitem[Ben{\'{\i}}tez et al.(2002)]{benitez2002} Ben{\'{\i}}tez, N., Ma{\'{\i}}z-Apell{\'a}niz, J., \& Canelles, M.\ 2002, Physical Review Letters, 88, 081101 
 
\bibitem[Binns et al.(2016)]{binns2016} Binns, W.~R., Israel, M.~H., Christian, E.~R., et al.\ 2016, Science, 352, 677 

\bibitem[Breitschwerdt et al.(2016)]{breitschwerdt2016} Breitschwerdt, D., Feige, J., Schulreich, M. M., et al. 2016,
  Nature, 532, 73

\bibitem[Capitanio et al.(2017)]{capitanio2017} Capitanio, L., Lallement, R., Vergely, J. L., et al. 2017, Astron.~Astrophys., 606, A65

\bibitem[Clayton \& Nittler(2004)]{clayton2004} Clayton, D.~D., \& Nittler, L.~R.\ 2004, \araa, 42, 39

\bibitem[Cowan et al.(2019)]{cowan2019} Cowan, J.~J., Sneden, C., Lawler, J.~E., et al.\ 2019, Reviews of Modern Physics, submitted, arXiv:1901.01410 

\bibitem[Cummings et al.(2002)]{cummings2002} Cummings, A.~C., Stone, E.~C., \& Steenberg, C.~D.\ 2002, \apj, 578, 194

\bibitem[Diehl(2018)]{diehl2018} Diehl, R.\ 2018, Astrophysics \& Space Science Library, 453, 3   

\bibitem[Diehl~et~al.(2006)]{diehl2006} Diehl, R., Halloin, H., Kretschmer, K., et al.\ 2006, \nat, 439, 45

\bibitem[Ellis and Schramm(1995)]{ellis+schramm1995} Ellis, J. \& Schramm, D.N. \ 1995, Proc. Nat. Acad. Sci., 92. 235

\bibitem[Ellis et al.(1996)]{ellis1996} Ellis, J., Fields, B.~D., \& Schramm, D.~N.\ 1996, \apj, 470, 1227

\bibitem[Faestermann(2018)]{faestermann2018} Faestermann, T.\ 2018, American Institute of Physics Conference Series, 1976, 020001   

\bibitem[Feige (2014)]{feige2014} Feige, J. %``Supernova-Produced Radionuclides in Deep-Sea Sediments Measured with AMS" 
  Doctoral Dissertation, University of Vienna, 2014 http://othes.univie.ac.at/35089

\bibitem[Feige et al.(2018)]{feige2018} Feige, J., Wallner, A., Altmeyer, R., et al.\ 2018, Physical Review Letters, 121, 221103

\bibitem[Fields et al.(2008)]{fields2008} Fields, B.~D., Athanassiadou, T., \& Johnson, S.~R.\ 2008, \apj, 678, 549 

\bibitem[Fimiani et al.(2016)]{fimiani2016} Fimiani, L., Cook, D.~L., Faestermann, T., et al.\ 2016, Physical Review Letters, 116, 151104

\bibitem[Fitoussi et al.(2008)]{fitoussi2008} Fitoussi, C., Raisbeck, G.~M., Knie, K., et al.\ 2008, Physical Review Letters, 101, 121101

\bibitem[Frisch(1981)]{frisch1981} Frisch, P.~C.\ 1981, \nat, 293, 377 

\bibitem[Frisch(2003)]{frisch1993} Frisch, P.~C.\ 1993, \apj, 407, 198 

\bibitem[Frisch \& Dwarkadas(2017)]{frisch2017} Frisch, P., \& Dwarkadas, V.~V.\ 2017, Handbook of Supernovae, 2253

\bibitem[Frisch et al.(2011)]{frisch2011} Frisch, P.~C., Redfield, S., \& Slavin, J.~D.\ 2011, \araa, 49, 237 
  
\bibitem[Fry et al.(2015)]{fry2015} Fry, B.~J., Fields, B.~D., \& Ellis, J.~R.\ 2015, \apj, 800, 71 

\bibitem[Fry et al.(2016)]{fry2016} Fry, B.~J., Fields, B.~D., \& Ellis, J.~R.\ 2016, \apj, 827, 48

\bibitem[Fry et al.(2018)]{fry2018} Fry, B.~J., Fields, B.~D., \& Ellis, J.~R.\ 2018, arXiv:1801.06859

\bibitem[Fryer et al.(2019)]{fryer2019} Fryer, C.~L., Timmes, F., Hungerford, A.~L., et al.\ 2019, arXiv:1902.02915

\bibitem[Gall et al.(2011)]{gall2011} Gall, C., Hjorth, J., \& Andersen, A.~C.\ 2011, Astron.~\& Astrophys.~Review, 19, 43 

\bibitem[Gehrels et al.(2003)]{gehrels2003} Gehrels, N., Laird, C.~M., Jackman, C.~H., et al.\ 2003, \apj, 585, 1169

\bibitem[Gonzalez et al.(2001)]{gonzalez2001} Gonzalez, G., Brownlee, D., \& Ward, P.\ 2001, Icarus, 152, 185

\bibitem[Hotokezaka et al.(2015)]{hotokezaka2015} Hotokezaka, K., Piran, T., \& Paul, M.\ 2015, Nature Physics, 11, 1042   

\bibitem[Hyde \& Pecaut(2018)]{hyde2018} Hyde, M., \& Pecaut, M.~J.\ 2018, Astronomische Nachrichten, 339, 78 

\bibitem[Kachelrie{\ss} et al.(2015)]{kachelriess2015} Kachelrie{\ss}, M., Neronov, A., \& Semikoz, D.~V.\ 2015, Physical Review Letters, 115, 181103

\bibitem[Kachelrie{\ss} et al.(2018)]{kachelriess2018} Kachelrie{\ss}, M., Neronov, A., \& Semikoz, D.~V.\ 2018, \prd, 97, 63011.  

\bibitem[Knie et al.(1999)]{knie1999} Knie, K., Korschinek, G., Faestermann, T., et al.\ 1999, Physical Review Letters, 83, 18 

\bibitem[Knie et al.(2004)]{knie2004} Knie, K., Korschinek, G., Faestermann, T., et al.\ 2004, Physical Review Letters, 93, 171103

\bibitem[Korschinek(2017)]{korschinek2017} Korschinek, G.\ 2017, Handbook of Supernovae, 2419 

\bibitem[Korschinek et al.(1996)]{Korschinek1996} Korschinek, G., Faestermann, T., Knie, K. \& Schmidt, C. 1996,
  Radiocarbon, 38, 68

\bibitem[Limongi \& Chieffi(2006)]{limongi2006} Limongi, M., \& Chieffi, A.\ 2006, \apj, 647, 483

\bibitem[Lineweaver et al.(2004)]{lineweaver2004} Lineweaver, C.~H., Fenner, Y., \& Gibson, B.~K.\ 2004, Science, 303, 59

\bibitem[Looney et al.(2006)]{looney2006} Looney, L.~W., Tobin, J.~J., \& Fields, B.~D.\ 2006, \apj, 652, 1755   

\bibitem[Ludwig et al.(2016)]{ludwig2016} Ludwig, P., Bishop, S., Egli, R., et al.\ 2016, Proceedings of the National Academy of Science, 113, 9232

\bibitem[Lugaro et al.(2018)]{lugaro2018} Lugaro, M., Ott, U., \& Kereszturi, {\'A}.\ 2018, Progress in Particle \& Nuclear Physics, 102, 1 

\bibitem[Mamajek(2016)]{mamajek2016} Mamajek, E.~E.\ 2016, Young Stars \& Planets Near the Sun, 314, 21 

\bibitem[McComas et al.(2009)]{mccomas2009} McComas, D.~J., Allegrini, F., Bochsler, P., et al.\ 2009, Science, 326, 959 

\bibitem[Melott \& Thomas(2019)]{melott2019a} Melott, A.~L., \& Thomas, B.~C.\ 2019, Journal of Geology in press, arXiv:1903.01501   

\bibitem[Melott et al.(2017)]{melott2017} Melott, A.~L., Thomas, B.~C., Kachelrie{\ss}, M., Semikoz, D.~V., \& Overholt, A.~C.\ 2017, \apj, 840, 105

\bibitem[Melott \& Thomas(2018)]{melott2018} Melott, A.~L., \& Thomas, B.~C.\ 2018, Lethaia, 51, 325, arXiv:1712.02730 

\bibitem[Melott et al.(2019)]{melott2019b} Melott, A.~L., Marinho, F., \& Paulucci, L.\ 2019, Astrobiology in press, arXiv:1712.09367

\bibitem[Meyer \& Clayton(2000)]{meyer2000} Meyer, B.~S., \& Clayton, D.~D.\ 2000, \ssr, 92, 133
  
\bibitem[Morrison \& Gowanlock(2015)]{morrison2015} Morrison, I.~S., \& Gowanlock, M.~G.\ 2015, Astrobiology, 15, 683   

\bibitem[M{\"u}ller et al.(2006)]{mueller2006} M{\"u}ller, H.-R., Frisch, P.~C., Florinski, V., \& Zank, G.~P.\ 2006, \apj, 647, 1491 

\bibitem[Nozawa et al.(2007)]{nozawa2007} Nozawa, T., Kozasa, T., Habe, A., et al.\ 2007, \apj, 666, 955 

\bibitem[Ruderman(1974)]{ruderman1974} Ruderman, M.A. \ 1974, Science 184, 1079

\bibitem[Savchenko et al.(2015)]{savchenko2015} Savchenko, V., Kachelrie{\ss}, M., \& Semikoz, D. V. 2015, \apj, 809,
L23

\bibitem[Schindewolf(1954)]{schindewolf1954} Schindewolf, O.H. \ 1954, Neues Jarhrbuch f\"ur Geologie und Paleont\"aontologie Monatshefte, 10, 457

\bibitem[Schulreich et al.(2017)]{schulreich2017} Schulreich, M. M., Breitschwerdt, D., Feige, J., et al. 2017, Astron.~Astrophys., 604, A81

\bibitem[Schulreich et al.(2018)]{schulreich2018} Schulreich, M. M., Breitschwerdt, D., Feige, J., et al. 2018, Galaxies, 6, 26  

\bibitem[Slavin \& Frisch(2008)]{slavin2008} Slavin, J.~D., \& Frisch, P.~C.\ 2008, \aap, 491, 53 

\bibitem[S{\o}rensen et al.(2017)]{sorensen2017} S{\o}rensen, M., Svensmark, H., \& Gr{\aa}e J{\o}rgensen, U.\ 2017, arXiv:1708.08248 

\bibitem[Sukhbold et al.(2016)]{sukhbold2016} Sukhbold, T., Ertl, T., Woosley, S.~E., Brown, J.~M., \& Janka, H.-T.\ 2016, \apj, 821, 38

\bibitem[Tetzlaff et al.(2013)]{tetzlaff2013} Tetzlaff, N., Torres, G., Neuh\"auser, R. \& Hohle, M.~M. 2013, MNRAS, 435, 879   
  
\bibitem[Thomas et al.(2016)]{thomas2016} Thomas, B.~C., Engler, E.~E., Kachelrie{\ss}, M., et al.\ 2016, \apjl, 826, L3
  
\bibitem[Tur et al.(2010)]{tur2010} Tur, C., Heger, A., \& Austin, S.~M.\ 2010, \apj, 718, 357 

\bibitem[Wallner et al.(2015)]{wallner2015} Wallner, A., Faestermann, T., Feige, J., et al.\ 2015, Nature Communications, 6, 5956   

\bibitem[Wallner et al.(2016)]{wallner2016} Wallner, A., Feige, J., Kinoshita, N., et al.\ 2016, \nat, 532, 69

\bibitem[Wang et al.(2007)]{wang2007} Wang, W., Harris, M.~J., Diehl, R., et al.\ 2007, \aap, 469, 1005
  
\end{thebibliography}
\end{document}